\documentclass{nature}
\usepackage{epsfig}
\usepackage{amsmath}

\usepackage{amsmath,mathrsfs,amsbsy,color,graphicx,bm,amsthm,amsfonts}
\usepackage{units}
\usepackage{bbm}
\usepackage{times}
\usepackage{dcolumn}
\usepackage{mathrsfs}% Align table columns on decimal point
\usepackage{amsmath,amssymb,epsfig,float}
\begin{document}
\baselineskip=0.6 cm
\title{Relativistic Quantum Metrology in Open System Dynamics}
\author{Zehua Tian$^{1}$, Jieci Wang$^{1,2,\dag}$, Heng Fan$^{2}$, and Jiliang Jing$^{1,\star}$}
\maketitle

\begin{abstract}
\baselineskip=0.5 cm
Quantum metrology studies
the ultimate limit of precision in estimating a physical quantity
if quantum strategies are exploited. Here we investigate the evolution of a two-level atom as a detector which interacts with a massless scalar field using the master equation approach for open quantum system. We employ local quantum estimation theory to estimate the Unruh temperature when probed by a uniformly accelerated detector in the Minkowski vacuum. In particular, we evaluate the Fisher information (FI) for population measurement, maximize its value over all possible detector preparations and evolution times, and compare its behavior with that of the quantum Fisher information (QFI). We find that the optimal precision of estimation is achieved when the detector evolves for a long enough time. Furthermore, we find that in this case the FI for population measurement is independent of initial preparations of the detector and is exactly equal to the QFI, which means that population measurement is optimal. This result demonstrates that the achievement of the ultimate bound of precision imposed by quantum mechanics is possible. Finally, we note that the same configuration is also available to the maximum of the QFI itself.
\end{abstract}

\begin{affiliations}
\item
Department of Physics, and Key Laboratory of Low
Dimensional, Quantum Structures and Quantum
Control of Ministry of Education,
 Hunan Normal University, Changsha, Hunan 410081, China.
\item
Beijing National Laboratory for Condensed Matter Physics, Institute
of Physics, Chinese Academy of Sciences, Beijing 100190, China.

$^\dag$e-mail: jieciwang@gmail.com
\\
$^\star$Corresponding author, e-mail: jljing@hunnu.edu.cn

\end{affiliations}

It is well known that in the modern theory of quantum fields, the concept of particle is observer-dependent \cite{Birell}. One of the most fundamental manifestations of this fact is the Unruh effect \cite{Birell,Unruh}, i.e., the inertial vacuum is perceived by a uniformly accelerated observer as populated by a thermal bath of radiation. It is believed that this effect is deeply connected with important physical phenomena such as Hawking radiation \cite{Hawking1,Hawking2,Kamil1,Kamil2}. Thus, its observation would be expected to provide experimental support for Hawking radiation and black hole evaporation. Furthermore, the detection of the Unruh effect would have an immediate impact in many fields such as astrophysics \cite{Astro,Alexandreas}, cosmology \cite{Cosm}, black hole physics \cite{BH}, particle physics \cite{PP}, quantum gravity \cite{QG} and relativistic quantum information \cite{RQI1,RQI2}. However, although a large number of different schemes involving Bose-Einstein condensates \cite{Retzker,Oren,Jaskula,Boiron} and superconducting circuits \cite{Nation,Wilson} have been proposed to detect the associated radiation effect, it remains an open research program to detect this effect in experiments, this is because the associated temperature lies far below any observable threshold (smaller than 1 Kelvin even for accelerations as high as $10^{21}m/s^2$). Since the Unruh effect is rather weak, high-precision quantum measurement is essential during its detection. On the other hand, due to the fact that nature is both quantum and relativistic, it can be expected by theoretical arguments that the Unruh effect is incorporated into the question of how to process information by using quantum technologies which are beyond the classical approaches \cite{RQI1,RQI2}. This creative combination provides not only a more complete frame to understand the theory of quantum information but also perhaps a new way to address the problem of ``information loss" in black hole scenarios. In particular, within this area at the overlap of relativity and quantum mechanics, it seems natural to apply novel approaches and techniques for quantum measurements. This makes the relativistic aspect of the effects potentially more accessible to detection.

The Unruh temperature of interest to us is nonlinear function of the density matrix and cannot, even in principle, correspond to a proper quantum observable. Therefore, its direct observation is not accessible. In these situations one has to turn to indirect measurements, inferring the value of the quantity of interest by inspecting a set of data coming from the measurement of a different observable, or a set of observables. In this regard, let us note that any conceivable strategy aimed at evaluating the quantity of interest ultimately reduces to a parameter-estimation problem that may be properly addressed in the framework of quantum estimation theory (QET) \cite{Helstrom,Holevo,Braunstein,Giovannetti1,Giovannetti2,Paris}. Relevant examples of this situation are given by discussions of quantum speed limits in open system dynamics \cite{Campo,Taddei,Deffner},  measurements of non-Markovianity of open quantum processes \cite{Lu}, estimation of quantum phase \cite{Giovannetti1,Giovannetti2,Monras,Dorner,Humphreys}, qubit thermometry \cite{Brunelli1,Brunelli2}, and so on. For example, with the help of rigorous methods from quantum statistics and estimation \cite{Helstrom}, recently Aspachs \emph{et al.} \cite{Aspachs} have investigated the ultimate precision limits for the estimation of the Unruh-Hawking temperature. Shorter after that, a number of analogous papers have emerged to study the topic of the estimation of relativistic effects \cite{Aspachs,Ahmadi1,Ahmadi2,Wang,Sabin,Jason,Doukas}.

Up to date, almost all work involving relativistic metrology is guided by an interesting link between field theory and quantum information: The change of coordinates between an inertial observer and a noninertial observer in the description of the state of a scalar field is equivalent to the transformation that affects a light beam undergoing parametric down-conversion in an optic parametric oscillator \cite{Birell, Walls}. The parameters encoded in quantum fields are assumed to be directly estimated without any scheme that investigates how to extract this information from the fields (relevant processes involve to how to introduce a probe and prepare what kind of probe). Besides, the quantum states to probe relativistic effects are directly prepared with the free field mode \cite{Aspachs,Wang,Sabin,Jason,Doukas}, which, as we all known, is spatially not localized and thus cannot be experimentally accessed and measured by localized apparatuses \cite{Dragan}.

Motivated by these considerations, in this work we employ a uniformly accelerated and localized two-level atom as the probe to detect the Unruh temperature. We aim at estimating the inverse Unruh temperature $\beta=1/T$ and try to address the following questions: (1) Which is the best probe state? (2) Which is the optimal measurement that should be performed at the output probe state? (3) Which is the minimum fluctuation in the temperature estimation, as well as the ultimate bound to precision imposed by quantum mechanics.

\section*{Results}

\subsection{Physical model and probe state preparation.}

We consider a two-level atom as the detector which interacts with a fluctuating vacuum scalar field. This model assumes that the detector behaves like an open system, i.e., a system immersed in an external field. Therefore, in the following we will treat the detector as an open quantum system and the vacuum with the fluctuations of the quantum field as the environment.

Let us first introduce the total Hamiltonian of the total system, detector plus field. Without loss of generality, it is taken as \begin{eqnarray}\label{Hamiltonian}
H=H_s+H_{\Phi(x)}+H_I,
\end{eqnarray}
where $H_s=\frac{1}{2}\omega_0\sigma_z$ and $H_{\Phi(x)}$ are respectively the Hamiltonian of the detector and scalar field, and $H_I=\mu(\sigma_++\sigma_-)\Phi(x(\tau))$ represents their interaction. Note that $\omega_0$ is the detector's energy-level spacing, $\sigma_z$ is the Pauli matrix, $\sigma_+$ ($\sigma_-$) is the atomic rasing (lowering) operator, and $\Phi(x)$ corresponds to the scalar field operator. Here, the two-level atom can be fully described in terms of a two-dimensional Hilbert space. Its quantum state, with respect to a fixed and arbitrary basis in this space, will be represented by a $2\times2$ density matrix $\rho$, which is Hermitian $\rho^\dagger=\rho$, and normalized $\mathrm{Tr}(\rho)=1$ with $\det(\rho)\geq0$. On the other hand, the equation of motion of the scalar field is $(\Box+m^2)\Phi=0$ with $\Box\Phi=g^{\mu\nu}\nabla_\mu\nabla_\nu\Phi=(-g)^{-1/2}\partial_\mu[(-g)^{-1/2}g^{\mu\nu}\partial_\nu\Phi]$, where $m$ is the mass of the field, and $g$ is the determinant of the metric $g_{\mu\nu}$. For the Minkowski spacetime case, one set of solutions of this equation of motion is $u_\mathbf{k}(t,\mathbf{x})=[2\omega(2\pi)^3]^{-1/2}e^{i\mathbf{k}\cdot\mathbf{x}-i\omega t}$ with $\omega=(\mathbf{k}^2+m^2)^{1/2}$. The field modes $u_\mathbf{k}$ and their respective complex conjugates form a complete orthonormal basis, so $\Phi$ may be expanded as $\Phi(x)=\sum_\mathbf{k}[a_\mathbf{k}u_\mathbf{k}(t,\mathbf{x})+a^\dagger_\mathbf{k}u^\ast_\mathbf{k}(t,\mathbf{x})]$, which plays a crucial role in the following calculation of the two point function of quantum field.

Initially, the total quantum system is described by the density matrix $\rho_{tot}=\rho(0)\otimes|-\rangle\langle-|$, in which $\rho(0)$ is the reduced density matrix of the detector, and $|-\rangle$ is the vacuum of the field defined by $a_\mathbf{k}|-\rangle=0$ for all $\mathbf{k}$. In the frame of the detector, the evolution in the proper time $\tau$ of the total density matrix $\rho_{tot}$ satisfies
\begin{eqnarray}\label{evolution equation}
\frac{\partial\rho_{tot}(\tau)}{\partial\tau}=-iL_H[\rho_{tot}(\tau)],
\end{eqnarray}
where the symbol $L_H$ represents the Liouville operator associated with $H$, $L_H[S]=[H,S]$. To obtain the dynamics of the detector, we must trace over the field degrees of freedom. After that, in the limit of weak coupling the evolving density matrix $\rho(\tau)$ of the detector obeys an equation in the Lindblad form \cite{Lindblad1,Lindblad2,Benatti1}
\begin{eqnarray}\label{Lindblad equation}
\frac{\partial\rho(\tau)}{\partial\tau}&=&-i[H_{eff},\rho(\tau)]+\mathcal{L}[\rho(\tau)]
\end{eqnarray}
with\begin{eqnarray}\label{Effective H}
\nonumber
&&H_{eff}=\frac{1}{2}\Omega\sigma_z=\frac{1}{2}\{\omega_0+\mu^2\mathrm{Im}(\Gamma_++\Gamma_-)\}\sigma_z,
\\
&&\mathcal{L}[\rho(\tau)]=\sum^3_{j=1}[2L_j\rho L^\dagger_j-L^\dagger_jL_j\rho-\rho L^\dagger_jL_j],
\end{eqnarray}
where
$\Gamma_\pm=\int^{\infty}_{0}e^{i\omega_0s}G^+(s\pm i\epsilon)ds$,
$L_1=\sqrt{\frac{\gamma_-}{2}}\sigma_-$, $L_2=\sqrt{\frac{\gamma_+}{2}}\sigma_+$, $L_3=\sqrt{\frac{\gamma_z}{2}}\sigma_z$,
$\gamma_\pm=2\mu^2\mathrm{Re}\Gamma_\pm$,
$\gamma_z=0$,
$G^+(x-x')=\langle0|\Phi(x)\Phi(x')|0\rangle$ is the field correlation function, and $s=\tau-\tau'$. Eq. (\ref{Lindblad equation}) characterizes the evolution of the detector. In particular, the second on its right hand side denotes
the dissipation resulting from the external environment, i.e., the scalar field that the detector couples to. It is called the Lindblad term and describes the response of the detector to the environment. All the information that we are interested in and want to estimate in the following is encoded in its relevant parameters.

In order to solve the Eq. (\ref{Lindblad equation}), let us express the reduced density matrix in terms of the Pauli matrices,
\begin{eqnarray}\label{PM}
\rho(\tau)=\frac{1}{2}\left(1+\sum^3_{i=1}\rho_i(\tau)\sigma_i\right).
\end{eqnarray}
If we choose the initial state of the detector as $|\psi(0)\rangle=\sin\frac{\theta}{2}|0\rangle+e^{-i\phi}\cos\frac{\theta}{2}|1\rangle$, substituting Eq. (\ref{PM}) into (\ref{Lindblad equation}), we can obtain its analytical evolving matrix,
\begin{eqnarray}\label{matrix for atom}
\rho(\tau)=\frac{1}{2}
\left(
\begin{array}{cc}
\rho_{ee}(\tau) & \rho_{eg}(\tau) \\
\rho_{ge}(\tau) & \rho_{gg}(\tau)
\end{array}
\right)
\end{eqnarray}
with
\begin{eqnarray}\label{elements}
\nonumber
\rho_{ee}(\tau)&=&1+e^{-A\tau}\cos\theta+\frac{B}{A}(1-e^{-A\tau}),
\\ \nonumber
\rho_{gg}(\tau)&=&1-e^{-A\tau}\cos\theta-\frac{B}{A}(1-e^{-A\tau}),
\\
\rho_{eg}(\tau)&=&\rho^\ast_{ge}(\tau)=e^{-\frac{1}{2}A\tau-i(\Omega\tau+\phi)}\sin\theta,
\end{eqnarray}
where $A=\gamma_++\gamma_-$ and $B=\gamma_+-\gamma_-$. Moreover, the state of the detector can be diagonalized and decomposed as  $\rho(\tau)=\lambda_+|\psi_+(\tau)\rangle\langle\psi_+(\tau)\rangle|+\lambda_-|\psi_-(\tau)\rangle\langle\psi_-(\tau)\rangle|$ with
\begin{eqnarray}\label{eigenvalues}
\nonumber
\lambda_\pm&=&\frac{1}{2}(1\pm\eta),
\\
|\psi_\pm(\tau)\rangle&=&\frac{[|\rho_{eg}(\tau)||0\rangle+e^{-i(\Omega\tau+\phi)}(\rho_{ee}(\tau)-2\lambda_\mp)|1\rangle]}
{\sqrt{(\rho_{ee}(\tau)-2\lambda_\mp)^2+|\rho_{eg}(\tau)|^2}},
\end{eqnarray}
where $\eta=\sqrt{(\rho_{ee}(\tau)-1)^2+|\rho_{eg}(\tau)|^2}$.

From Eqs. (\ref{Lindblad equation}) and (\ref{Effective H}) we know that the Wightman function for the scalar field that the detector couples to plays an important role in the evolution of the detector. In this regard, let us note that if
a uniformly accelerated detector with trajectory, $t(\tau)=\frac{1}{a}\sinh(a\tau),~x(\tau)=\frac{1}{a}\cosh(a\tau),~y(\tau)=z(\tau)=0$,
is coupled to a massless scalar field in the Minkowski vacuum, then the corresponding Wightman function should be \cite{Birell}
\begin{eqnarray}
G^+(x,x')=-\frac{a^2}{16\pi^2}\sinh^{-2}\bigg[\frac{a(\tau-\tau')}{2}-i\varepsilon\bigg].
\end{eqnarray}
In this case, it is easy to obtain
\begin{eqnarray}\label{AB}
A=\frac{\mu^2\omega_0}{2\pi}\bigg(\frac{e^{2\pi\omega_0/a}+1}{e^{2\pi\omega_0/a}-1}\bigg),~~~
B=-\frac{\mu^2\omega_0}{2\pi}.
\end{eqnarray}

Substituting Eq. (\ref{AB}) into (\ref{matrix for atom}), it is easy to check that when evolving long enough time, i.e., $\tau\gg\frac{1}{\gamma_++\gamma_-}$ with $\frac{1}{\gamma_++\gamma_-}$ being the time scale for atomic transition, the detector eventually approaches to the state
\begin{eqnarray}\label{thermal}
\rho(\infty)=\frac{e^{-\beta H_s}}{Tr[e^{-\beta H_s}]}.
\end{eqnarray}
Here let us remark that the state in Eq. (\ref{thermal}) is a thermal state with a temperature $T=1/\beta$.  Thus, the accelerated detector feels as if it were immersed in a thermal bath with temperature $T=a/2\pi$ \cite{Birell}. We will estimate this relativistic parameter in the following.

\subsection{Fisher information based on population measurement.}
As we stated in the Discussion section, the QFI determines the ultimate bound on the precision of the estimator although it is then difficult to find out which measurement is optimal to achieve such ultimate bound. This occurs because the QFI does not depend on any measurements, for it is obtained by maximizing the FI over all possible quantum measurements on the quantum system. Thus, to find out the optimal measurement to estimate the Unruh temperature, we first calculate the FI for the population measurement, and then compare the FI with the QFI to determine whether the population measurement is optimal according to the condition of optimal quantum measurement, i.e., POVM with a FI equal to the QFI. For the population measurement, the FI, according to Eqs. (\ref{matrix for atom}) and (\ref{FI}), is given by
\begin{eqnarray}\label{PMFI}
\nonumber
F(\beta)&=&\frac{[\partial_\beta p(e|\beta)]^2}{p(e|\beta)}+\frac{[\partial_\beta p(g|\beta)]^2}{p(g|\beta)}
\\
&=&\frac{1}{2}\bigg[\frac{[\partial_\beta\rho_{ee}(\tau)]^2}{\rho_{ee}(\tau)}
+\frac{[\partial_\beta\rho_{gg}(\tau)]^2}{\rho_{gg}(\tau)}\bigg].
\end{eqnarray}
Substituting Eqs. (\ref{elements}) and (\ref{AB}) into Eq. (\ref{PMFI}), we can obtain the detailed formula of the FI. It is interesting to note that the FI is independent of quantum phase $\phi$. It only depends on the parameters $\tau, \theta$ and $\omega_0$. Thus, the FI in fact should be written as $F(\beta,\tau,\theta,\omega_0)$, while we adopt the notation $F(\beta)$ for convenience here. In the following, by evaluating the FI we want to find both the optimal initial detector preparation and the smallest temperature value that can be discriminated. We will work with dimensionless quantities by rescaling time and temperature
\begin{eqnarray}\label{RSTT}
\tau\longmapsto\widetilde{\tau}\equiv\gamma_0\tau,~~~~\beta\longmapsto\widetilde{\beta}\equiv\beta\omega_0,
\end{eqnarray}
where $\gamma_0=\frac{\mu^2\omega_0}{2\pi}$ is the spontaneous emission rate of the atom. For convenience, we continue to term $\widetilde{\beta}$ and $\widetilde{\tau}$, respectively, as $\beta$ and $\tau$.

Let us consider that the detector is uniformly accelerated with proper acceleration $a$ and Unruh temperature $T$ proportional to $a$ \cite{Unruh}. We assume that the inverse temperature has the value $\beta=10$. The probabilities $p(j|\beta)=\rho_{jj}(\tau)$ evolve according to Eq. (\ref{elements}). The corresponding behavior of the FI is shown in  the top panel of Fig. 1. We can see that for $\theta=\pi$ the FI is larger than the FI of other cases during initial period, but when $\tau\gg\frac{1}{\gamma_++\gamma_-}$ all the FI are saturated and equal to each other. This means that the FI displays a robust maximum at the optimal time $\tau\gg\frac{1}{\gamma_++\gamma_-}$ for all $\theta$. We can also obtain the same results from the bottom panel of Fig. 1. It is shown that the FI evolves periodically as a function of the initial state parameter $\theta$ and for any time the maximal FI is always obtained by taking $\theta_{\max}=\pi$, i.e., by preparing the detector in the ground state. Furthermore, for small time the FI suddenly drops to zero, except for a sharp peak centered at $\theta_{\max}$, as $\theta$ varies, but for long time the FI changes less with respect to $\theta$. Thus, we can arrive at the conclusion that the maximum sensitivity in the predictions for the inverse Unruh temperature can be obtained by initially preparing the detector in its ground state. However, if the detector evolves for a long enough time, the maximum sensitivity in the predictions is independent on the initial state in which the detector is prepared. It is no surprise because the accelerated detector eventually evolves to a thermal state regardless of its initial state \cite{Benatti2}.

In Fig. 2 we plot the FI for different fixed temperatures as a function of time $\tau$. We can see that the FI approaches its maximum value when the detector evolves for a long enough time. Also its value for different $\beta$ varies over several orders of magnitude, changing from $10^{-5}$ to $0.1$, which means the FI is very dependent on the temperature itself. As we demonstrated above, the reason for saturation is that the accelerated detector eventually evolves to a thermal state regardless of its initial state \cite{Benatti2}. Furthermore, in this case, the thermal state only depends on the thermal temperature felt by the detector, i.e., the acceleration of the detector. On the other hand, the higher the temperature, the bigger the FI is, i.e., the easier it is to achieve a given precision in the estimation of temperature.

\subsection{ Quantum Fisher information}

In order to assess the performance of the population measurement in the estimation of the Unruh temperature we have evaluated the QFI of the family of states $\rho(\tau)$ in Eq. (\ref{matrix for atom}). Substituting Eqs. (\ref{eigenvalues}) and (\ref{AB}) into (\ref{QFIH}), it is easy to obtain the QFI. Let us note that the QFI depends on $\beta$, $\tau$ and $\theta$, but is independent of the phase $\phi$ of the detector. Thus, to find out the optimal working regimes we have to maximize the value of the QFI over all three relevant parameters.

Similar to the analyses of the FI shown above, we first fix the Unruh temperature by assuming $\beta=10$, and discuss how the effective time $\tau$ (initial state parameter $\theta$) affects the QFI for different initial state parameters $\theta$ (effective time $\tau$). Obviously, from Fig. 3 we know that the maximum of the QFI is achieved by initially preparing the detector in the ground state. However, if the effective time is long enough, i.e., the detector evolves for a long enough time, $\tau\gg\frac{1}{\gamma_++\gamma_-}$, no matter what the initial state is prepared in, the QFI always achieves the maximum, which means the optimal sensitivity in estimation of $\beta$ is independent of the initial preparation of the detector if the effective time is long enough. Besides, in Fig. 4 we plot the QFI for different fixed temperatures as a function of the effective time $\tau$. We find that, if the detector evolves for a long enough time, the QFI we computed above for different Unruh temperatures saturates at different values which vary over several orders of magnitude. Furthermore, the higher the temperature, the bigger the QFI is, i.e., the easier it is to achieve a given precision in the estimation of temperature. Thus, we can arrive at the conclusion that the maximum sensitivity in the predictions for the inverse Unruh temperature can be obtained when the detector evolves for a long enough time, and the maximum sensitivity in the predictions is independent on the initial state in which the detector is prepared. In this case we want to emphasize that this strategy provides optimality in the sense that inequality (\ref{QCR}) is saturated and the variance Var($\beta$) is as small as possible.

We find that for $\tau\gg\frac{1}{\gamma_++\gamma_-}$ both the FI and QFI take the maximum limit. Interestingly, upon inspecting the temporal evolution of the excited state probability, $p(e|\beta)$ has a minimum under this condition (also the quantum state of the detector is thermal discussed in Eq. (\ref{thermal})). Thus, we can give a physical explanation to the FI and QFI behavior. Because we want to estimate a tiny quantity that carries information about thermal disorder, of course, only when the external environment is mostly occupied by the Unruh thermal particle, and the more the better, we then could expect to find the maximum sensitivity in the predictions. This condition corresponds to the probability $p(e|\beta)$ achieving its minimum.

In the above analysis, we have shown the behaviors of the FI and QFI, and obtained the conditions that how to achieve the maximum FI and QFI. It is interesting to note that the behavior of $H(\beta)$ is identical to that of $F(\beta)$, as is apparent by comparing Fig. 1 and 3. Besides, under the same condition $(\theta, \phi, \tau)=(\theta, \phi, \infty)$ both the FI and QFI obtain the maximum value when $\beta$ is fixed. In order to find out whether the population measurement is optimal during the estimation process of the Unruh temperature, we will check whether the maximized FI is equal to the optimal QFI. Thus, we prepare the detector in its ground state, i.e., $\theta=\pi$, and assume that the detector evolves for a long enough time. This allows us to easily find that the detector eventually evolves to a thermal state. In this case, the off-diagonal terms of state (\ref{matrix for atom}) vanish and it is diagonal with two eigenvalues
\begin{eqnarray}\label{thermal state}
\nonumber
\lambda_+&=&\frac{1}{e^{2\pi\omega_0/a}+1},
\\
\lambda_-&=&\frac{e^{2\pi\omega_0/a}}{e^{2\pi\omega_0/a}+1},
\end{eqnarray}
and corresponding eigenvectors $|e\rangle$ and $|g\rangle$. For this quantum statistic model, we find that the FI is equal to the QFI given by
\begin{eqnarray}\label{FH}
F(\beta)=H(\beta)=\frac{(\partial_\beta\lambda_+)^2}{\lambda_+}+\frac{(\partial_\beta\lambda_-)^2}{\lambda_-}.
\end{eqnarray}
It means that the estimation of $\beta$ via the population measurement is optimal. Eq. (\ref{FH}) is the ultimate bound to precision of estimation of the Unruh temperature. Because the population measurement is optimal, our results in this regard suggest that the achievement of the ultimate bound to precision of estimation of the Unruh temperature allowed by quantum mechanics is in the capability of current technology.

\section*{Discussion}

We introduced a detector, i.e., a two-level atom, which is uniformly accelerated and interacts with a massless scalar field in the Minkowski vacuum, and employ it to detect the Unruh temperature. By employing local quantum estimation theory we have studied the estimation of the Unruh temperature via quantum-limited measurements performed on the detector. In particular, we have analyzed the precision of estimation as a function of both the detector initial preparations and the interaction parameters, and evaluated the limits of precision posed by quantum mechanics.

It is shown that the FI for the population measurement, which establishes a classical bound on precision, takes the maximum limit when the detector evolves for a long enough time compared with the time scale for atomic transition, $\frac{1}{\gamma_++\gamma_-}$, i.e., when $\tau\gg\frac{1}{\gamma_++\gamma_-}$. In this case, the FI for population measurement is independent of any initial preparations of the detector. Furthermore, we find that the same configuration is also corresponding to the maximum of the QFI based on all possible quantum measurements, which establishes the ultimate bound to the precision allowed by quantum mechanics. Interestingly, the maximum FI is equal to the maximum QFI under the same conditions, which means the optimal measurement for the estimation of the Unruh temperature corresponds to the population measurement. Thus, during the detection of the Unruh temperature, we can achieve the ultimate bound to the precision by performing a population measurement on the detector, and the ultimate bound is given by Eq. (\ref{FH}). Because the population measurement is allowed by the current technology \cite{Leibfried1,Leibfried2,Leibfried3,Freyberger,Bardroff,Arnold,Dada}, our results, in this regard, indicate that the ultimate bound to precision of estimation of Unruh temperature imposed by quantum mechanics can in principle be achieved under the current technology. On the other hand, our results demonstrate that thermalized quantum statistic model, Eq. (\ref{thermal}), plays an optimal role in the estimation of the Unruh temperature. This occurs because we want to estimate a tiny quantity that carries information about thermal disorder. Therefore, it is natural to expect to find the maximum sensitivity in the predictions when the external environment, that is coupled with the detector, is mostly occupied by the Unruh thermal particle, and the more the better. This condition corresponds to when $\tau\gg\frac{1}{\gamma_++\gamma_-}$, i.e., when the detector state is thermalized.

Our model avoids two  critical technical difficulties in the estimation of the Unruh temperature: a physically unfeasible detection of global free mode in the full space \cite{Aspachs} and a non-analytical expression of QFI due to the boundary conditions of the moving cavity \cite{Ahmadi1,Ahmadi2}. Recently, the open quantum system approach has been used to understand the Hawking effect of black hole \cite{Yu1} and Gibbons-Hawking effect of de sitter universe \cite{Yu2}. Thus, our above analysis can also be applied to discussing the estimation of Hawking temperature and Gibbons-Hawking temperature. Also we could turn to the estimation of other parameters, such as the atomic frequency and phase, analyzing what kind of role that the relativistic effects play in this metrology. In particular, the simulation of relativistically accelerating atoms in trapped ion systems and superconducting circuits has been studied in Ref. \cite{martin}. The simulations proposed in Ref. \cite{martin} are precise analogues of the physical setting required here. Our techniques could possibly be implemented during such simulations.

\begin{methods}

Usually, two main steps are contained in estimation process: at first we has to choose a measurement, and then, after collecting a sample of outcomes, we should find an estimator, i.e., a function to process data and to infer the value of the quantity of interest.
For a given measurement scheme, the mean square error $\mathrm{Var}(\beta)=E_\beta[(\hat{\beta}-\beta)^2]$ of any estimator of the parameter, $\beta$, is bounded by the Cram\'{e}r-Rao inequality \cite{Helstrom}
\begin{eqnarray}\label{QCR inequlity}
\mathrm{Var}(\beta)\geq\frac{1}{MF(\beta)},
\end{eqnarray}
where $M$ is the number of identical measurements repeated and $F(\beta)$ is the FI given by
\begin{eqnarray}\label{FI}
F(\beta)=\sum_jp(j|\beta)(\partial_\beta\ln p(j|\beta))^2=\sum_j\frac{|\partial_\beta p(j|\beta)|^2}{p(j|\beta)}.
\end{eqnarray}
Efficient estimators are those saturating the Cram\'{e}r-Rao inequality. In order to obtain the ultimate bound to precision, i.e., the smallest value of the parameter that can be discriminated, the optimization of FI is needed via a suitable choice of all its dependent parameters. From Eqs. (\ref{elements}) and (\ref{FI}), the FI obviously depends on the detector initial state parameters and evolving time, and so on. In this regard, let us note that the initial states of the detector and evolving time play an important role in this metrology process, which essentially determine the ultimate bound on precision.

On the other hand, we can also maximize the FI over all possible quantum measurements on the quantum system. By introducing the symmetric Logarithmic Derivative (SLD) satisfying $\frac{L_\beta\rho_\beta+\rho_\beta L_\beta}{2}=\frac{\partial\rho_\beta}{\partial\beta}$, the FI of any quantum measurement is upper bounded by the so-called QFI given by
\begin{eqnarray}\label{QCR}
F(\beta)\leq H(\beta)=Tr\big(\rho_\beta L^2_\beta\big).
\end{eqnarray}
Here, it is interesting to note that the QFI does not depend on any measurements carried on the detector, indeed being obtained by maximizing over all possible measurements \cite{Paris}. Further studies show that the detailed formula for the QFI is of \cite{Paris}
\begin{eqnarray}\label{QFIH}
H(\beta)=\sum_{k=\pm}\frac{(\partial_\beta\lambda_k)^2}{\lambda_k}
+2\sum_{k\neq k'=\pm}\frac{(\lambda_k-\lambda_{k'})^2}{\lambda_k+\lambda_{k'}}
\bigg|\langle\psi_k|\partial_\beta\psi_{k'}\rangle\bigg|^2,
\end{eqnarray}
where $\lambda_k$ and $|\psi_{k'}\rangle$ satisfy $\rho_\beta=\sum_k\lambda_k|\psi_k\rangle\langle\psi_k|$. The first term in Eq. (\ref{QFIH}) represents the classical Fisher information whereas the second term contains the truly quantum contribution. Therefore, it is natural to generalize the Cram\'{e}r-Rao inequality (\ref{QCR inequlity}) to its quantum version
\begin{eqnarray}\label{QCR1}
\mathrm{Var}(\beta)\geq\frac{1}{MH(\beta)},
\end{eqnarray}
which shows the ultimate bound to the precision allowed by quantum mechanics for a given statistical model $\rho(\beta)$.

For a given quantum measurement, i.e., a POVM, Eq. (\ref{FI}) establishes the classical bound on precision, which may be achieved by a proper data processing, e.g., by maximum likelihood, which is known to provide an asymptotically efficient estimator. On the other hand, Eq. (\ref{QFIH}) establishes the ultimate bound to the precision allowed by quantum mechanics. Thus optimal quantum measurement for the estimation of $\beta$ corresponds to POVM with a FI equal to the QFI, i.e., those saturating inequality (\ref{QCR}). In our paper, we calculate the FI for the population measurement, i.e., $|e\rangle\langle e|$ and $|g\rangle\langle g|=\mathbf{1}-|e\rangle\langle e|$ with outcomes probabilities $\{Tr[\rho_\beta|e\rangle\langle e|], Tr[\rho_\beta|g\rangle\langle g|]\}$, performed on the detector, and maximize it over all the parameters it depends on. Then we compare it with the QFI based on all possible quantum measurements to find out whether the population measurement is optimal and determinate the ultimate bound to the precision.

\end{methods}

\begin{addendum}

\item [Acknowledgement]

This work is supported by the  National Natural Science Foundation
of China under Grant Nos. 11175065, 11475061, 11305058; the National Basic
Research of China under Grant No. 2010CB833004; the SRFDP under
Grant No. 20114306110003.

\item [Author Contributions]
Z. T. made the main calculations.
 J. J., J. W.,and H. F. discussed the results, Z. T. wrote the paper with assistances from J. J. and other authors.

\item [Competing Interests]
The authors declare that they have no competing financial interests.

\item [Correspondence]
Correspondence and requests for materials should be addressed to
J. J.
\end{addendum}

\newpage

\textbf{Figure 1. By taking $\beta=10$ the FI is plotted as a function of the effective time $\tau$ with different $\theta$ values.}
(The top one) $\theta=\pi$ (dot-dashed red line), $\theta=0.95\pi$ (dashed blue line), $\theta=0$ (solid green line). (The bottom one) the FI is plotted for $\beta=10$ as a function of the initial state parameter $\theta$ with different effective time $\tau$: $\tau=10$ (dot-dashed red line), $\tau=5$ (dashed blue line), $\tau=1$ (solid green line).

\textbf{Figure 2. Log-linear plot of the FI as a function of effective time $\tau$ with different values of $\beta$.}
The detector is initially prepared in its ground state $|0\rangle$ ($\theta=\pi$). From bottom to top, $\beta=10$ (dot-dashed red line), $\beta=6$ (dashed blue line), $\beta=2$ (solid green line).

\textbf{Figure 3.  By taking $\beta=10$ the QFI is plotted as a function of the effective time $\tau$ with different $\theta$ values.}
(The top one) $\theta=\pi$ (dot-dashed red line), $\theta=0.95\pi$ (dashed blue line), $\theta=0$ (solid green line). (The bottom one) the QFI is plotted for $\beta=10$ as a function of the initial state parameter $\theta$ with different effective time $\tau$: $\tau=10$ (dot-dashed red line), $\tau=5$ (dashed blue line), $\tau=1$ (solid green line).

\textbf{Figure 4. Log-linear plot of the QFI as a function of effective time $\tau$ with different values of $\beta$. }
The detector is initially prepared in its ground state $|0\rangle$ ($\theta=\pi$). From bottom to top, $\beta=10$ (dot-dashed red line), $\beta=6$ (dashed blue line), $\beta=2$ (solid green line).

\newpage
\begin{figure}
\begin{center}
\epsfig{file=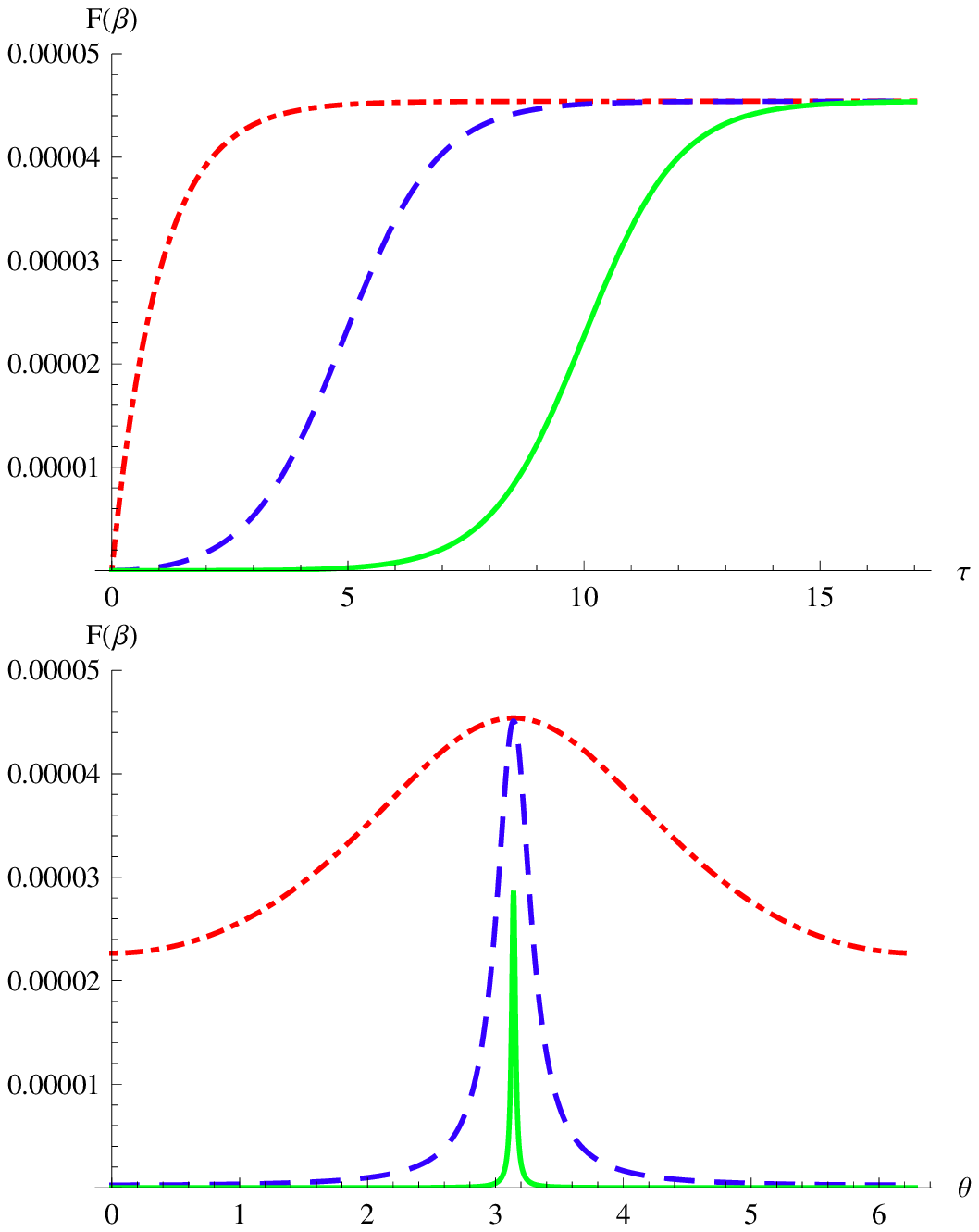,width=8cm}
\end{center}
\par
%\caption{}
\label{F1}
\end{figure}

\begin{center}
Figure1
\end{center}

\newpage
\begin{figure}
\begin{center}
\epsfig{file=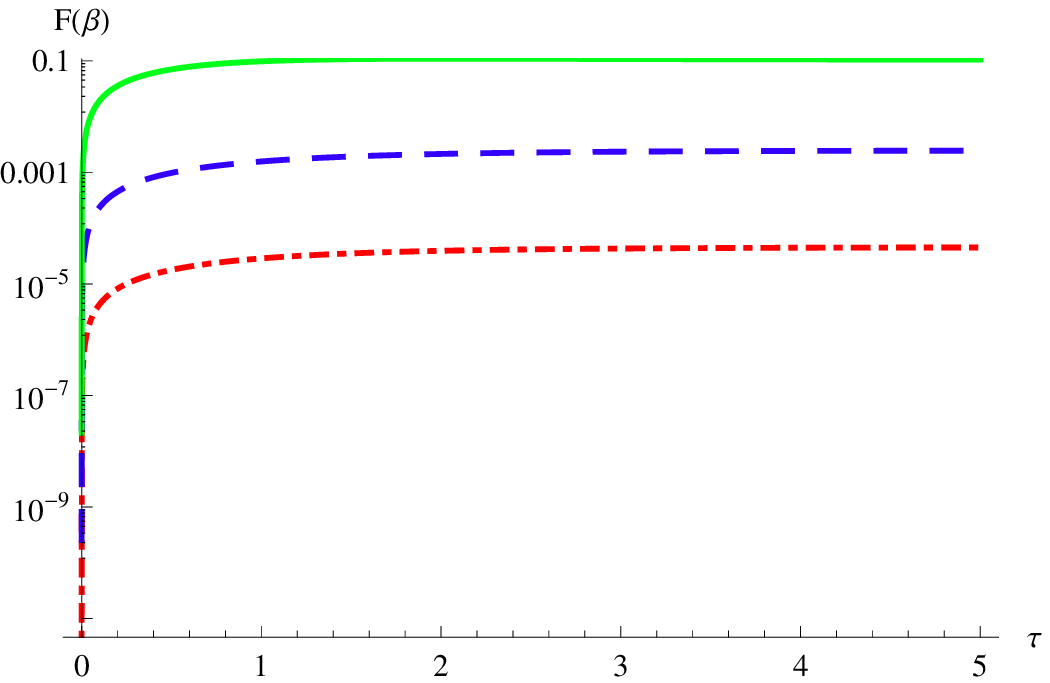,width=8cm}
\end{center}
\par
%\caption{}
\label{Fig2}
\end{figure}

\begin{center}
Figure2
\end{center}

\newpage
\begin{figure}
\begin{center}
\epsfig{file=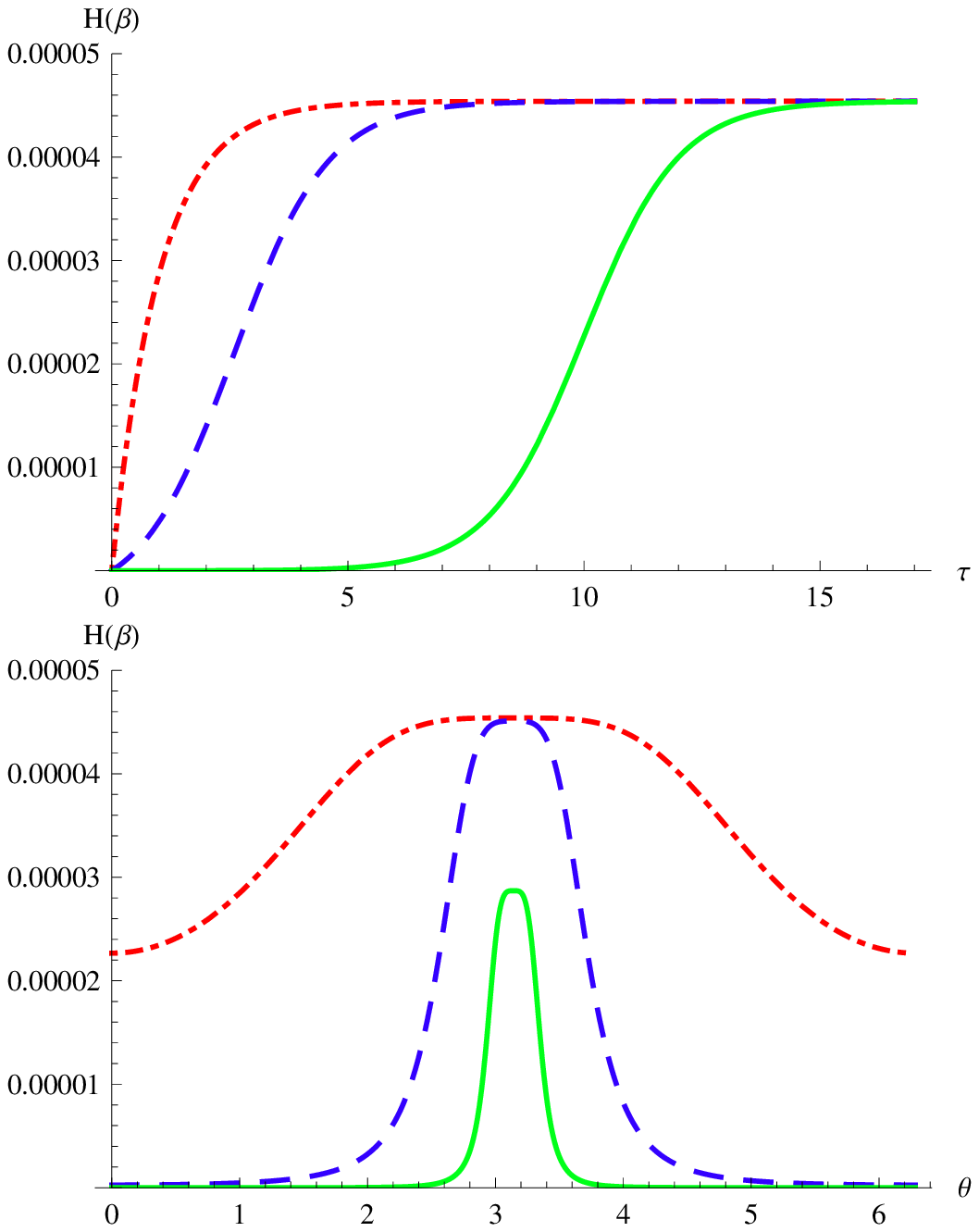,width=8cm}
\end{center}
\par
%\caption{}
\label{Fig3}
\end{figure}

\begin{center}
Figure3
\end{center}

\newpage
\begin{figure}
\begin{center}
\epsfig{file=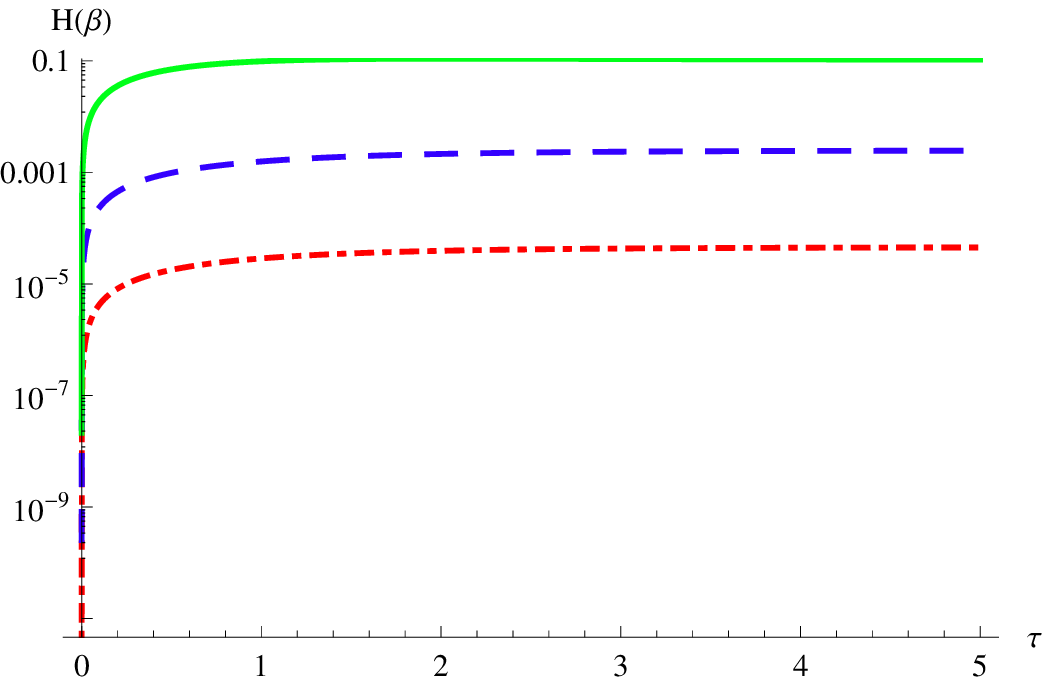, width=8cm}
\end{center}
\par
\label{Fig4}
\end{figure}

\begin{center}
Figure4
\end{center}

\newpage

\end{document}